\documentstyle[epsfig]{elsart}
\begin{document}
\begin{frontmatter}
\title{\bf Energy calibration of the NaI(Tl) calorimeter of
           the SND detector using cosmic muons }
\author{ M.N.Achasov\thanksref{adrr},}
\author{ A.D.Bukin, }
\author{ D.A.Bukin, }
\author{ V.P.Druzhinin, }
\author{ V.B.Golubev, }     
\author{ V.N.Ivanchenko,}
\author{ S.V.Koshuba, }     
\author{ S.I.Serednyakov } 
\thanks[adrr]{ E-mail: achasov@inp.nsk.su, FAX: +7(383-2)35-21-63}
\address{ Institute of Nuclear Physics,
          Novosibirsk,
          630090,
          Russia }
\date{}
\begin{abstract}
 The general purpose spherical nonmagnetic detector ( SND ) is now
 taking data at VEPP-2M $e^+e^-$ collider in BINP ( Novosibirsk )
 in the centre of mass energy range of $0.2 \div 1.4$ GeV.
 The energy calibration of the NaI(Tl)
 calorimeter of the SND detector with cosmic muons is
 described. Using this method, the energy resolution of
 $5.5~\%~(\sigma)$ for 500 MeV photons was achieved. 
\end{abstract}
\end{frontmatter}

\section{Introduction}
 Electromagnetic scintillation calorimeters are an important part of many
 elementary particle detectors. One of the most complicated
 problems of such calorimeters is the energy calibration, i.e. determination
 of coefficients, needed for conversion of electrical signals
 from the calorimeter into corresponding energy depositions, measured in
 units of MeV.  Commonly used for that purpose
 are particles, producing known energy depositions in the calorimeter
 counters, e.g. gamma quanta from radioactive sources, cosmic muons,
 or particle beams. 

 Cosmic radiation is a continuos and freely available source of
 charged particles. At the sea level the main part of them is represented by
 muons $(\sim 75 \% )$, having high enough energies
 $( \sim 1 \mathrm{GeV})$ to traverse the whole detector. In the calorimeter
 counters muons loose energy due to ionization
 of the medium. The average ionization losses are close to
 5 MeV/cm in NaI(Tl). With reasonable accuracy one can also assume,
 that the trajectory of the muon is a straight line and that energy is
 left mainly in the counters crossed by this line. 

 Cosmic calibration of the SND calorimeter is a preliminary stage
 before the final calibration. Its goal is to obtain  approximate values of
 the constants for the calculation of energy depositions. These
 constants are needed 
 to level responses of all crystals in order to obtain equal 
 first level trigger energy threshold over the whole calorimeter.
 They are also used as
 seed values for a precise OFF-LINE calibration procedure,
 based on analysis of Bhabha events.
 The requirements for the cosmic calibration procedure are the
 following: the calibration procedure must take not more than 12
 hours, it must be independent of the other detector
 systems, and statistical errors in the conversion coefficients
 must be less than $1\%$. 

 The scheme of this procedure was briefly described in \cite{nim}.
 Its important feature is that it makes use of virtually all
 muons detected in the calorimeter, greatly reducing the time required for
 calibration. In many other implementations of cosmic
 calibration, e.g. in the L3 detector at LEP \cite{L3}, muons,
 traversing the detector only in certain directions, were selected.

\section{The SND calorimeter}

\begin{figure}
\epsfig{figure=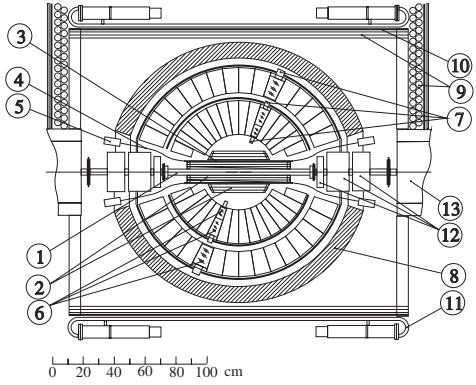,height=5cm}
\caption{SND detector, section along the beams; 1 --- beam pipe,
         2 --- drift chambers, 3 --- scintillation counters,
         4 --- light guides, 5 --- PMTs, 6 --- NaI(Tl) crystals,
         7 --- vacuum phototriodes, 8 --- iron absorber,
         9 --- streamer tubes, 10 --- 1cm iron plates,
         11 --- scintillation counters, 12 and 13 --- 
         elements of collider magnetic system.}
\label{f1}
\end{figure}

 The SND detector \cite{snd1,snd2} (Fig.~\ref{f1}) consists of two
 cylindrical drift chambers
 for charge particle tracking, calorimeter, and a muon system.

\begin{figure}
\epsfig{figure=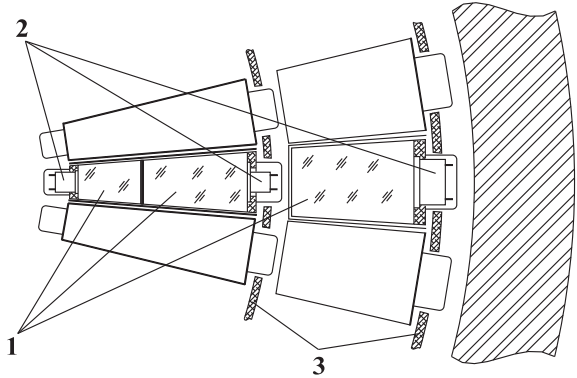,height=5cm}
\caption{NaI(Tl) crystals layout inside the calorimeter:
         1 --- NaI(Tl) crystals, 2 --- vacuum phototriodes,
         3 --- aluminum supporting hemispheres.}
\label{f4}
\end{figure}

 The main part of the detector is a three-layer spherical calorimeter,
 based on NaI(Tl) crystals. The pairs of counters of the two inner layers, 
 with thickness of 2.9
 and 4.8 $X_0$ respectively, where $X_0=2.6~\mathrm{cm}$ is a radiation length,
 are sealed in thin aluminum containers, fixed
 to an aluminum supporting hemisphere. Behind it, the third layer of 
 NaI(Tl) crystals, 5.7 $X_0$ thick, is placed (Fig.~\ref{f4}).
 The total thickness of the calorimeter for particles, originating from
 the center of the detector, is equal to 34.7 cm (13.4 $X_0$).
 The total number of counters is 1632, the number of crystals per
 layer varies from 520 to 560. The angular dimensions of the most of
 crystals are $\Delta\phi = \Delta\theta=9^\circ$, the total solid
 angle is $90\%$ of $4\pi$.

\begin{figure}
\epsfig{figure=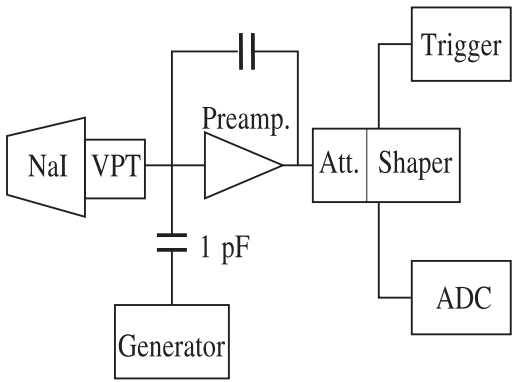,height=5cm}
\caption{Electronics channel of the SND calorimeter:
         NaI --- NaI(Tl) scintillator, VPT --- vacuum phototriode,
         Preamp --- charge sensitive preamplifier,
         Att --- computer controlled attenuator, Shaper --- shaping amplifier,
         Generator --- calibration generator,
         ADC --- analog to digital converter, Trigger --- first level trigger.}
\label{f5}
\end{figure}

 The electronics channel (Fig.~\ref{f5}) consists of:
\begin{enumerate}
\item
 phototriode with an average quantum efficiency of the photocathode for the
 NaI(Tl) emission spectrum of about $15\%$ and gain of about 10 \cite{b2}.
 The light collection efficiency varies from 7 to 15 $\%$ for different
 calorimeter layers,
\item
 charge sensitive preamplifier (CSA) with a conversion coefficient of~
 0.7~V/pC,
\item
 12-channel shaping amplifier (SHA) with a remote controlled gain that
 can be set to any value in the range from 0 to a maximum with a resolution
 of 1/255.
\item
 24-channel 12 bit analog to digital converter (ADC) with a maximum
 input signal $U_{\mathrm{max}}=2 \mathrm{V}$,
\item
 in addition the SHA produces special signals for the first level trigger,
 the most important of which is the analog sum of all calorimeter channels
 --- the total energy deposition.
\end{enumerate}

 Each  calorimeter channel
 can be tested using a precision computer controlled calibration generator.
 The amplitude of its signal can be set to any value from 0 to 1 V 
 with a resolution of
 1/4096. The equivalent electronics noise of individual calorimeter
 counters lies within a range of 150-350 keV.

\section{Algorithm of the detector calibration with cosmic muons}
 The calibration procedure is based on the comparison of experimental
 energy depositions in the calorimeter crystals for cosmic muons with
 Monte Carlo simulation. The simulation was carried out by means of the
 UNIMOD2 \cite{b3} code, developed for simulation of $e^+e^-$ colliding beam
 experiments. Simulated were muons crossing the sphere with a radius of
 80 cm, centered at the beam crossing point, and containing all calorimeter
 crystals. The cosmic muon generation program is based on data from \cite{b4}.
 Experimental data on the energy and angular spectra of cosmic muons were
 approximated using the following formula:
\begin{equation}
 {\partial N \over \partial p \partial \theta }
 =F_{0} \cdot \cos^{k}\theta\cdot\exp \biggl[-{ \ln^2(p/p_{0}) \over
 2\sigma^2}\biggr],
\end{equation}
 where $F_0=3\cdot10^{-6}{\mathrm{cm}}^{-2}{\mathrm{s}}^{-1}
 {\mathrm{ster}}^{-1}{\mathrm{(MeV/c)}}^{-1}$ --- the flux of muons with
 a momentum of $p=p_0=500\mathrm{MeV/c}$ at $\theta=0$,
 $\theta$ --- angle between the muon momentum and vertical direction,
 $k=2.86$, $\sigma=(2.86-\cos\theta))/1.54$.
 The data from \cite{b4}, on the relative abundances of positive and negative
 muons, were approximated using the following expression:
\begin{equation}
 N_+/N_-=1,0+0,35\cdot \exp \biggl[-0,22 \ln^2\biggl({p \over
 7\mathrm{GeV/c}}\biggr)\biggr].
\end{equation}
 The passage of particles through the detector was simulated by means of the
 UNIMOD2 code. As a result the energy depositions in the calorimeter crystals
 were calculated in each event. A total of 1 000 000 cosmic events were
 simulated.

 The simulated events were processed as follows. The track of cosmic muon was
 fitted to the hit crystals and expected track lengths
 in each crystal were calculated.
 Energy  depositions $E_{\mathrm{i}}$ and track lengths $l_{\mathrm{i}}$
 were summed over event sample for each crystal separately and for each
 crystal their ratio was calculated:
 $C_{\mathrm{mc}}(\mathrm{MeV/cm})= 
 \langle E(\mathrm{MeV})\rangle / \langle l(\mathrm{cm})\rangle$.
 Brackets denote averaging over event samples. Then
 the calorimeter channels were calibrated electronically, using the
 generator. In order to account for the electronics nonlinearity,
 the calibration data were taken at two significantly different values
 of SHA gain and wide range of calibration generator amplitudes, covering
 the whole dynamic range of the ADCs.
 In addition, data were taken at a working value of the SHA gain.

 Using all these data and assuming that the dependence of the ADC counts
 $S_{\mathrm{adc}}$ on the generator amplitude $U_{\mathrm{gen}}$ is
 described by a second order polynomial, and that the generator amplitude
 itself linearly depends on generator code, one can obtain
 $a,~b,~c,~p,~K_{\mathrm{u}}$ constants, defined by the following 
 expressions:
\begin{equation}
 S_{\mathrm{adc}} = aU_{\mathrm{gen}}^2  + bU_{\mathrm{gen}} + c~~~
 \mathrm{and}~~~U_{\mathrm{gen}} = K_{\mathrm{u}}( G + p ),
\end{equation}
 where $G$ and $p$ are a generator code and pedestal respectively,
 $K_{\mathrm{u}}$ - the relative SHA gain.
 The next step is to obtain from these intermediate results the
 final constants $A$, $B$, and $C$ for the working values of SHA gain,
 from the knowledge of which one can calculate the equivalent generator
 code $G$ for any measured ADC count:
\begin{equation}
 G = AS_{\mathrm{adc}}^2 + BS_{\mathrm{adc}} + C.
\end{equation}
 It relates an ADC count to a corresponding generator code,
 which allows the linearization of the ADC response.

\begin{figure}
\epsfig{figure=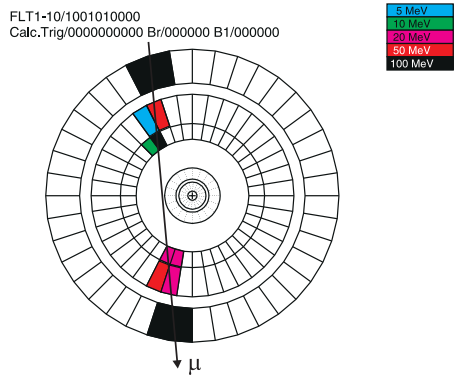,height=5cm}
\caption{Example of cosmics event selected by calibration procedure.}
\label{f9}
\end{figure}

 The next stage is processing of the experimental events.
 About 1.5 million events (Fig.~\ref{f9}) are needed to be collected in
 special data taking runs with total duration of 4.5 hours.
 The processing itself is the same as that
 of the simulated events, but now the initial ADC counts are first
 transformed into equivalent generator codes. As a result, values of
 $C_{\mathrm{exp}}(\mathrm{Gen/cm})= \langle G(\mathrm{Gen}) \rangle
 / \langle l(\mathrm{cm}) \rangle$  for each crystal are obtained.
 The meaning of $C_{\mathrm{exp}}$ is a code, which
 should be written into the generator to produce signals on CSA input,
 equivalent to an average input signal produced by a 1 cm long muon track.

 Using $C_{\mathrm{exp}}(\mathrm{Gen/cm})$ and
 $C_{\mathrm{mc}}(\mathrm{MeV/cm})$, for each crystal we can calculate 
\begin{equation}
 C_{\mathrm{cal}}(\mathrm{MeV/Gen}) = 
 C_{\mathrm{mc}}(\mathrm{MeV/cm})/C_{\mathrm{exp}}(\mathrm{Gen/cm})
 ~~~ \mbox{---}
\end{equation}
 an equivalent generator code, corresponding to an energy deposition of 1 MeV
 in a crystal. Then the gains of the SHA channels are adjusted to equalize
 contributions of different crystals into the total energy deposition
 signal and the final generator calibration pass is carried out.
 It yields the coefficients
 $a_{\mathrm{k}}$, $b_{\mathrm{k}}$, and $c_{\mathrm{k}}$,
 needed to transform $S_{\mathrm{adc}}$ into MeV according to the
 following formula:
\begin{equation}
 E(\mathrm{Mev})=a_{\mathrm{k}}S_{\mathrm{adc}}^2+b_{\mathrm{k}}
 S_{\mathrm{adc}}+c_{\mathrm{k}}.
\end{equation}

 The use of the precision generator as a reference calibration source
 is very convenient not only because its linearity
 is much higher than that of the ADC, but it also simplifies the replacement of
 any element of the electronics channel, except the CPA or generator itself.
 All what is needed after that is to recalibrate the calorimeter 
 with the generator.

\section{The description of the event processing algorithm}

 The goal of the experimental and simulated events processing is to obtain
 the normalized ratios $C_{\mathrm{mc}}(\mathrm{MeV/cm})$ and
 $C_{\mathrm{exp}}(\mathrm{Gen/cm})$. The need for normalization
 of the energy deposition to the unit track length in a crystal arises from the
 fact that the raw energy deposition spectra in crystals are very wide and
 have very weakly pronounced maxima. On the other hand, such a normalization
 reduces systematic errors due to possible differences in the angular
 distributions of
 simulated and real muons, and differences in the directions 
 of crystals axes over the calorimeter.
 The simulated and experimental events are processed using the same computer
 code. The
 only difference is that before processing of a simulated event it is checked
 for passing the experimental trigger.
 The following selection criteria are used for event selection:
 the total number of hit crystals must be from 5 to 25. Crystals with energy
 depositions smaller than the threshold value (currently 5 MeV) are discarded.
 Similarly discarded are crystals with ADC counts less than 20.
 An event is selected if more than four hit crystals survive these cuts.
 For those crystals a straight line is fitted, using a least squares method,
 i. e. the  sum of squares of distances from the crystal centers to the
 line is minimized. The line is parametrized  as:
\begin{equation}
 \vec{r}=\vec{R}+\vec{v}t,
\end{equation}
 where $\vec{R}$ is an arbitrary point on the line, $\vec{v}$ is a unitary
 vector in the direction of the line, $t$ --- parameter.
 Then, the squared distance from the center of i-th crystal to the line is:
\begin{equation}
 L_{\mathrm{i}}^2=(\vec{X_{\mathrm{i}}}-\vec{R})^{2}-
 (\vec{v}(\vec{X_{\mathrm{i}}}-\vec{R}))^{2}, 
\end{equation}
 where $\vec{X_{\mathrm{i}}}$ is a radius-vector of the crystal center.
 The minimized function is:
\begin{equation}
 F_{\mathrm{min}}(\vec{R},\vec{v})=
 \sum\limits_{\mathrm{i}}L_{\mathrm{i}}^2/\sigma_{\mathrm{i}}^{2},
\end{equation}
 where $\sigma_{\mathrm{i}}$ is assumed to be equal to a half height
 of an i-th crystal. One point on a line can be determined immediately:
\begin{equation}
 \vec{R_0}={\sum\limits_{\mathrm{i}}\vec{X_{\mathrm{i}}}/
 \sigma_{\mathrm{i}} \over
          \sum\limits_{\mathrm{i}}1/\sigma_{\mathrm{i}}^{2}}. 
\end{equation}
 The direction of vector $\vec{v}$, is determined by a maximum of the
 quadratic form $F(\vec{R_0},\vec{v})$.
\begin{equation}
 F(\vec{R_0},\vec{v})=
 \sum\limits_{\mathrm{i}}(\vec{v}(\vec{X_{\mathrm{i}}}
 -\vec{R_0}))^2/\sigma_{\mathrm{i}}^{2}.
\end{equation}
 In other words, $\vec{v}$ must have the same direction as an 
 eigenvector of $A$, matrix of this quadratic form, corresponding to its 
 maximal eigenvalue. Here 
\begin{equation}
 A_{\mathrm{mn}}=\sum\limits_{\mathrm{i}}
 ((X_{\mathrm{i}}^{\mathrm{m}}-R_{0}^{\mathrm{m}})
 (X_{\mathrm{i}}^{\mathrm{n}}-R_{0}^{\mathrm{n}})/
 \sigma_{\mathrm{i}}^{2}),
\end{equation}
 $\mathrm{m,~n} = 1, 2, 3$ --- coordinate indices. The equation for
 $\vec{v}$ was solved using an iterative procedure:
\begin{equation}
 \vec{v}_{\mathrm{n+1}}=A\vec{v}_{\mathrm{n}}/|A\vec{v}_{\mathrm{n}}|.
\end{equation}

\begin{figure}
\epsfig{figure=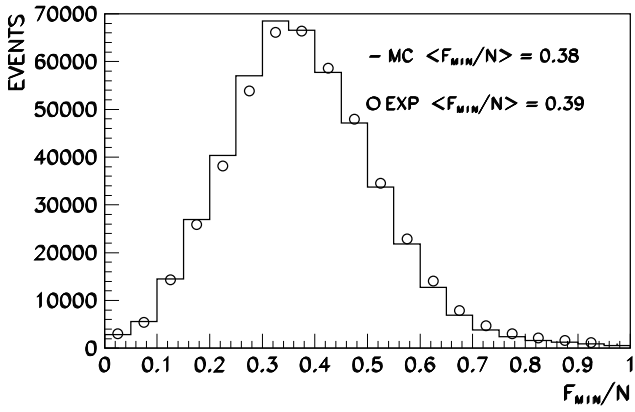,height=7cm}
\caption{Distribution over residual values of minimized function in 
         experimental and simulated events.}
\label{f7}
\end{figure}

 As seed value the unitary vector in the direction of the line 
 between the centers of the two most distant crystals in the event is used.
 On each 
 iteration step the presence of hit crystals at distances greater than 
 $2\sigma_{\mathrm{i}}$ from the line is checked. If such crystals exist,
 two options are considered. The first one is to remove the most distant 
 crystal from the list, the second one is to remove 
 the crystal, most distant from  the point $\vec{R_0}$. For both
 cases the fitting of the line is repeated and the final choice is based on the
 minimal value of $F_{\mathrm{min}}(\vec{R_0},\vec{v})$. If there are
 more than one crystal to be discarded then the event is rejected
 completely. From the remaining events only those are selected where
 $F_{\mathrm{min}}(\vec{R_0},\vec{v})$ divided by the number of hit
 crystals is less than 0.7 (Fig.~\ref{f7}).

 After that, the expected lengths of the muon track $l_{\mathrm{ij}}$ in each
 crystal are 
 calculated and averaged over the whole event sample together with the 
 energy depositions $E_{\mathrm{ij}}$. The values of average track
 lengths $\langle l_{\mathrm{i}} \rangle$ and  energy depositions
 $\langle E_{\mathrm{i}} \rangle$ are used then to calculate the energy 
 deposition per unit track length
 $C_{\mathrm{i}}=\langle E_{\mathrm{i}} \rangle /
 \langle l_{\mathrm{i}} \rangle$. Here $i$ and $j$ are the crystal and
 the event numbers, respectively.

 To estimate the statistical error in $C_{\mathrm{i}}$, the event sample is
 divided into groups of 50 events each and within each group the ratio 
 $R_{i}=\langle E_{i} \rangle / \langle l_{i} \rangle $ is calculated.
 Then, the error $\sigma_{\mathrm{Ri}}$ in $C_{\mathrm{i}}$ can be
 estimated as $\sigma_{\mathrm{Ri}}=(\langle R_{\mathrm{i}}^{2} \rangle -
 \langle R_{\mathrm{i}} \rangle^2)^{1/2}/\sqrt{N}$,
 where $N$ is a number of groups.

 The total CPU time, required for processing of $1.5 \cdot 10^6$ cosmic
 events is 2.5 hours on VAXstation~4000/60.

\section{Events processing results. Comparison of experimental and simulated
         distributions}

 The mean $\langle C_{\mathrm{mc}} \rangle$ and their RMS values are
 shown in the Table~\ref{tab1}. The necessity of normalization
 of energy depositions in crystals to a unitary track length was studied.
 To this end the events with muons going through drift chamber were processed,
 and the corresponding $C_{\mathrm{mc}}^{1}$ coefficients were calculated.
 The ratios of $\langle C_{\mathrm{mc}}/C_{\mathrm{mc}}^{1} \rangle$
 and their RMS are also presented in the Table~\ref{tab1} together with the
 corresponding ratios of mean energy depositions $\langle E/E^1 \rangle$.
 The data show, that the normalization significantly reduces the dependence of
 coefficients on the angular distribution of muons.

\begin{figure}
\epsfig{figure=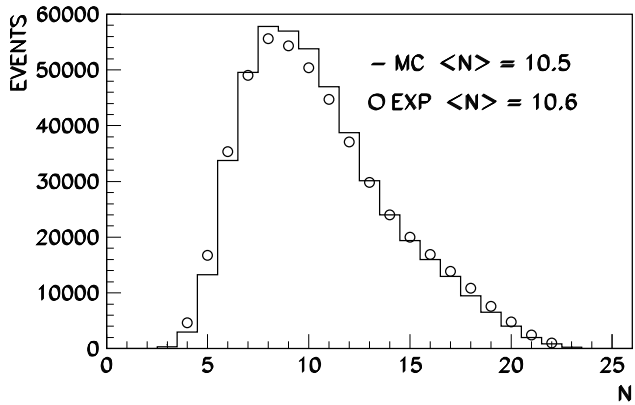,height=7cm}
\caption{Distributions over number of hit crystals for the events with energy
         depositions above certain threshold. N --- number of hit crystals.
         Circles --- experimental events, histogram --- MC simulation.}
\label{f6}
\end{figure}

\begin{figure}
\epsfig{figure=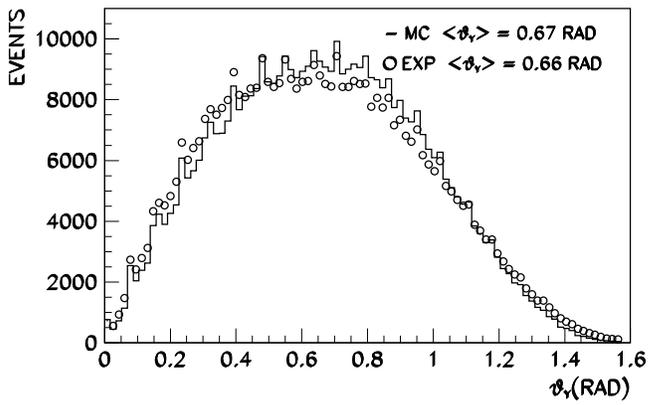,height=7cm}
\caption{Distribution over the track angle with respect to a vertical 
         direction.}
\label{f10}
\end{figure}

\begin{figure}
\epsfig{figure=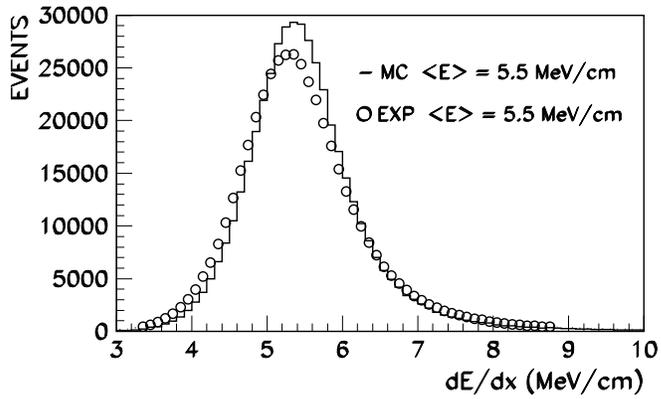,height=7cm}
\caption{Total energy depositions in the calorimeter per unit track 
         length in experimental and simulated events.}
\label{f12}
\end{figure}

 The distributions of experimental events over number of hit crystals with 
 energy depositions higher than threshold value and over 
 $F_{\mathrm{min}}(\vec{R_0},\vec{v})$ are in good agreement with the
 simulated ones (Fig.~\ref{f6},~\ref{f7}). Small differences between
 experimental and simulated distributions may  be attributed to inaccuracies
 in the angular and energy spectra of the primary particles and first level
 trigger
 simulation. This also may cause small differences in the distributions
 over the angle relative to  vertical direction
 (Fig.~\ref{f10}). Shown in Fig.~\ref{f12} are distributions over total 
 energy deposition in the detector, normalized to a unitary track length also 
 agree well.

 A comparison of average track lengths of muons in the calorimeter 
 crystals for the experimental and simulated events was carried out. The 
 results are listed in Table~\ref{tab2}. The statistical uncertainty of
 the $C_{\mathrm{exp}}$ coefficients is less than $1~\%$ for all three
 layers of the calorimeter. At the same time the uncertainties of the
 corresponding coefficients for simulated events are determined by
 simulation statistics.

\section{Energy resolution of the calorimeter. Implementation of the 
calibration procedure}

\begin{figure}
\epsfig{figure=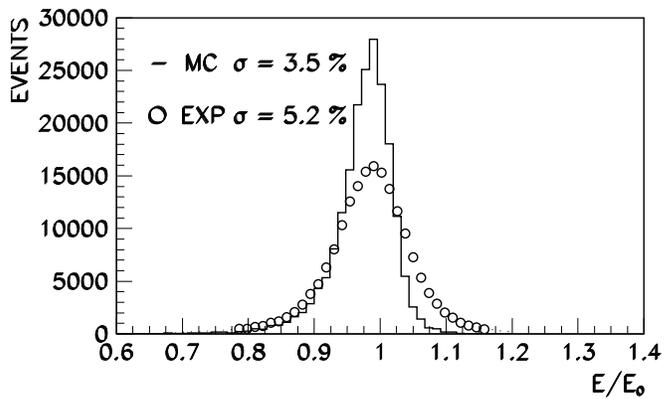,height=7cm}
\caption{Energy spectra for 500 MeV electrons. $E$ is a measured
         energy, $E_0$ is the beam energy.}
\label{f131}
\end{figure}

 The energy resolution of the calorimeter was studied using
 $e^+e^- \rightarrow e^+e^-$ and $e^+e^- \rightarrow \gamma\gamma$
 processes. The energy distributions for electrons are depicted in
 Fig.~\ref{f131}. The events with polar angles of the particles in the range
  $45 \leq \theta \leq 135$ degrees and  acollinearity angle 
 less than 10 degrees were selected. To estimate the energy resolution 
 quantitatively, the spectra in Fig.~\ref{f131} were fitted by a function
\begin{equation}
 F(E)=A \cdot\exp  \biggl\{-{1 \over 2} \cdot 
 \biggl[\biggl({\ln(\eta) \over t_{\mathrm{x}}}\biggr)^2+
 t_{\mathrm{x}}^2 \biggr] \biggr\}, 
\end{equation}
\begin{displaymath}
 \eta = 1+{t_{\mathrm{x}}(E-E_{\mathrm{m}}) \over \sigma} \cdot
 {\sinh (t_{\mathrm{x}} \sqrt[]{\ln 4}) \over t_{\mathrm{x}}\sqrt[]{\ln4}},
\end{displaymath}
 where $E$ is a particle energy, $E_{\mathrm{m}}$ --- the position of maximum,
 $A$ --- normalization coefficient, $t_{\mathrm{x}}$ - the asymmetry
 parameter, $\sigma$ is a full width of the distribution
 at half maximum divided by 2.36. $A$, $E_{\mathrm{m}}$, $\sigma$,
 and $t_{\mathrm{x}}$ were treated as free parameters during fitting.
 Energy resolution was defined as
 $\sigma_{\mathrm{E}} / E = \sigma / E_{\mathrm{m}}$.

 Experimental and Monte Carlo resolutions of the calorimeter at 500 MeV 
 are respectively $3.7~\%$ and $5.5~\%$ for photons and $3.5~\%$ and
 $5.2~\%$ for electrons. Peak and mean values of the experimental and
 simulated distributions agree to a level of about $1~\%$, but the widths of 
 experimental distributions are significantly larger. The possible 
 explanation of the broadening of the experimental spectra could be electronics 
 instability, systematic errors in calibration procedure and  nonuniformity 
 of light collection efficiency over the crystal volume.

 Study of the electronics and photodetector stability showed, that it can 
 attribute to a maximum of $1~\%$ shifts for the time duration
 of the collection of  energy deposition spectra.
 Systematic biases were estimated by 
 comparison of the calibration coefficients with those obtained using a 
 special procedure based on minimization of the width of the energy deposition 
 spectrum in the calorimeter for $e^+e^-$ events. Such a calibration was 
 carried out using experimental statistics collected in 1996 \cite{b5}. The 
 average bias of coefficients relatively to those obtained with cosmic 
 calibration was $3~\%$ for the first layer. No statistically significant 
 bias was found in other two layers. The RMS difference in 
 calibration coefficients, obtained using cosmic and $e^+e^-$ calibration 
 procedures is close to $3 \div 5~\%$. The resulting energy resolution after 
 $e^+e^-$ calibration was close to $5.0~\%$ for photons and $4.6~\%$ for 
 electrons, which is still worse than that expected from Monte Carlo 
 simulation.

\begin{figure}
\epsfig{figure=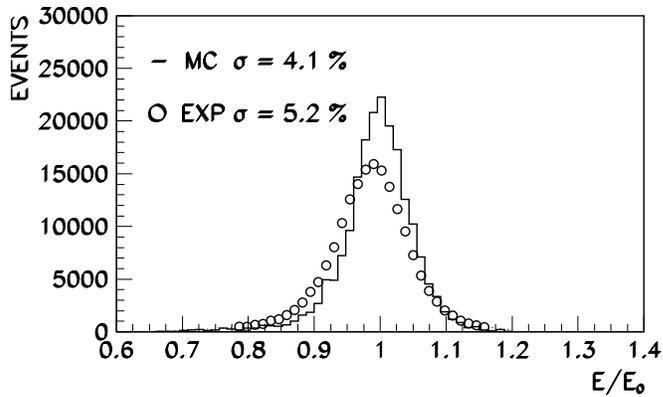,height=7cm}
\caption{Energy spectra for 500 MeV electrons. The nonuniformity
         of light collection efficiency over the crystal volume was taken
         into account in Monte Carlo simulation. $E$ is the electron measured
         energy, $E_0$ is the beam energy.}
\label{f132}
\end{figure}

 A Monte Carlo simulation was carried out taking into account a nonuniformity
 of light collection efficiency over
 the crystal volume. The results are shown
 in Fig.~\ref{f132} together with the experimental distribution.
 The energy resolution
 for simulated events decreased to $4.1~\%$. It shows that the broadening of
 experimental spectra was mainly determined by the nonuniformity of light
 collection efficiency over the crystal volume.

\begin{figure}
\epsfig{figure=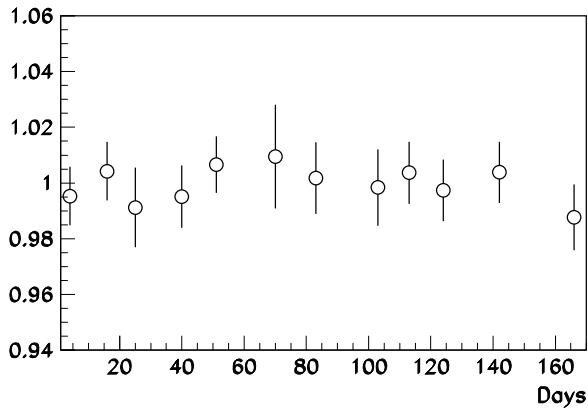,height=7cm}
\caption{The calibration coefficients spread. Points --- average ratio of 
         the current calibration results to the preceding ones,
         error bars --- FWHM of the distributions of these ratios over
         the whole calorimeter layer divided by 2.36. Horizontal axis shows
         the number of days elapsed from the first calibration. Shown
         are the results for the second layer. Other calorimeter layers behave
         similarly.}
\label{f14}
\end{figure}

 The cosmic calibration procedure was used during a five month experiment 
 with SND at VEPP-2M in 1996 \cite{b5}. Relative shifts in calibration 
 coefficients are shown in Fig.~\ref{f14}. It can be seen, that for a one week 
 period between consecutive calibrations, the mean shift of the coefficients 
 is less than $2~\%$ and RMS of their random spread is $1 \div 2~\%$.

\section{Conclusion}

With the help of the described procedure of SND calorimeter calibration 
a statistical accuracy of $1~\%$ in calibration coefficients and energy 
resolution of $5.5~\%$ for 500 MeV electromagnetic showers was achieved. 
Mean energy depositions agree with simulation at a level better than 
$1~\%$. The total time, required for calorimeter calibration is not more 
than 5 hours.

\newpage

\begin{table}
\caption{$C_{\mathrm{mc}}$ coefficients. Comparison of calibration
         results based on different muon samples (MC).}
\label{tab1}
\begin{tabular}[t]{|c|cc|cc|cc|}
\hline
 Layer number & $C_{\mathrm{mc}}$ & $\sigma_{\mathrm{C}}(\mathrm{MeV/cm})$ 
              & $\langle C_{\mathrm{mc}}/C_{\mathrm{mc}}^{1} \rangle$
              & $\sigma$
              & $\langle E/E^{1} \rangle$ & $\sigma$  \\ \hline
 I   & 5.38 & 0.14 &
       0.99 & 0.02 & 0.83 & 0.04 \\ \hline
 II  & 5.43 & 0.09 &
       0.96 & 0.03 & 0.71 & 0.05 \\ \hline
 III & 5.61 & 0.08 &
       0.99 & 0.03 & 0.68 & 0.05 \\ \hline
\end{tabular}
\end{table}

\begin{table}
\caption{Comparison of mean track lengths of muons in calorimeter
         crystals for the experimental and simulated events.
         The statistical uncertainty of $C_{\mathrm{mc}}$.}
\label{tab2}
\begin{tabular}[t]{|c|cc|c|}
\hline
 Layer number & $\langle \langle l_{\mathrm{mc}} \rangle /
                 \langle l_{\mathrm{exp}} \rangle \rangle$
              & $\sigma$ & $\sigma_{\mathrm{Ri}}/\sqrt[]{N}$ \\ \hline
 I   & 1.01 & 0.02 & $1.4\%$ \\ \hline
 II  & 1.00 & 0.01 & $1.0\%$ \\ \hline
 III & 1.00 & 0.01 & $0.9\%$ \\ \hline
\end{tabular}
\end{table}

\clearpage

\newpage

\end{document}